# Diffusion wave and signal transduction in biological live cells

Tian You Fan[1][*] and Lei Fan[2]

[1] Department of Physics, School of Physics, Beijing Institute of Technology, Beijing 100081, China

[2] Department of Biology, School of Life Science, Beijing Institute of Technology, Beijing 100081, China

[*] tyfan@bit.edu.cn; tyfan2006@yahoo.com.cn

**Abstract**

Transduction of mechanical stimuli into biochemical signals is a fundamental subject for cell physics. In the experiments of FRET signal in cells a wave propagation in nanoscope was observed. We here develop a diffusion wave concept and try to give an explanation to the experimental observation. The theoretical prediction is in good agreement to result of the experiment.

Mechanical effects of environment influence many essential processes of gene expression, cell survival and growth etc [1], among them the transduction of mechanical stimuli into biochemical signals is a fundamental subject for cell physics. However the mechanism on the transduction remains unsolved mystery so far [2-7].

Combination of fluorescent resonance energy transfer (FRET),

genetically encoded biosensor and laser-tweezer [8-11] makes a revolutionary progress of our understanding on the transduction above mentioned in the recent two decades. This experimental technology enables the imaging and quantification of spatio-temporal characterization of mechanotransduction to biochemical signals in live cells.

Wang et al. [9] carried out a systematical study by using fluorescent resonance energy transfer and developed a genetically encoded biosensor approach. They observed a series of signal transduction events at the cell membrane, in which the long-range wave propagation with speed $U = (18.1 \pm 1.7) nm/s$ along the cell membrane is especially interesting. They further measured the wave length in a certain extent, this provides the most important information for theoretical analysis.

Na et al. [2] summarized the relevant studies up to 2008 and developed the stress-induced Src activation, and analyzed some mechanism on the transduction, e.g. stress effect, interaction between integrins and cytoskeleton, deformation of mircotubules, effect of stiffness of substrates, conformational changes or unfolding of membrane-bound proteins, and diffusion of the proteins, and so on.

There is no possibility to explore all aspects concerning the mechanism, it is difficult even if revealing each among them. In this note we focus on the dynamic process explored by Wang et al. [9], and suggest a physical-mathematical model to explain the long-range wave propagation.

Experiment [9] is carried out in a fibronectin-integrin-cytoskeleton-cell memebrane of human umbilical vein endothelial cells (HUVEs), this is a complex of proteins. A local mechanical stimulation is implemented by the laser-tweezer traction (The experimental detail can be referred to Wang et al. [9]). This aroused a motion of the protein complex, and led to a kind of wave propagation of Src activation with new feature. This new feature should reflect the dynamic nature of interaction between proteins and cell membrane. Due to the unclearness of mechanism of the wave propagation, the following study has to undertake some phenomenological analyses.

***One-dimensional analysis***. Relative to slow wave propagation, the dimensional of cell body ( $1\mu m \times 1\mu m \times 1\mu m$ ) is quite large, we approximately assume that the cell membrane is a large plane for a certain time interval, we take the plane is $xy$–plane. Furthermore, for simplicity we consider the one-dimensional case first. If the protein concentration of a cell is denoted by $C(x,t)$, then the equation of motion of the membrane-bound proteins is

$$\frac{\partial C}{\partial t} = D \frac{\partial^2 C}{\partial x^2} \qquad (1)$$

in which $D$ represents diffusion coefficient of the membrane-bound proteins, $x$ the spatial coordinate, $t$ the time.

The local mechanical stimulation $f(x,t)$ (induced by external force $F(x,t)$, or displacement $u(x,t)$, or stress $\sigma(x,t)$) is not, in general, a periodic function, but which may be expanded by a Fourier series as follows

$$f(x,t) = \sum_{n=o}^{\infty} A_n e^{i(xk_n - \omega_n t + \alpha_n)}$$

in which each Fourier component is a periodic function. For simplicity, we consider a periodic mechanical stimulation. Equation (1) is diffusion equation, for the periodic stimulation, it can have a wave propagation solution for long-range wave propagation such as

$$C(x,t) = C_0 e^{i(kx - \omega t)} \qquad (2)$$

the wave is propagating along the cytoplasm membrane, in which $C_0$ is a constant, $k$ the wave number, $\omega$ the circular frequency (or angular frequency), $i = \sqrt{-1}$, respectively, and the wave speed $U$ is unknown at moment.

Substituting (2) into (1) yields

$$k = \frac{1}{\sqrt{2}} \sqrt{\frac{\omega}{D}} \qquad (3)$$

This is a new relation—dispersion relation between $\omega$ and $k$, in which the wave number is connected to diffusion coefficient $D$, this is quite different from those based on traditional wave theory. The wave described by (2) and (3) may be viewed as diffusion wave. According to the well-known wave law one can find the wave speed as

$$U = \frac{\partial \omega}{\partial k} \tag{4}$$

and substituting (3) into (4), we determine the wave speed as follows

$$U = 4kD = 8\pi \frac{D}{\lambda} \tag{5}$$

where $\lambda$ is the wave length and there is relation between the wave length $\lambda$ and the wave number $k$ as below

$$\lambda = \frac{2\pi}{k} \tag{6}$$

Formula (5) is also a new result, in which the wave speed depends upon not only quantity concerning wave (e.g. wave length) but also diffusion coefficient.

With the above theoretical results, we can analysis on experimental phenomena. As reported in Ref [9], the observed values of wave length for the signal propagation are

$$\lambda = 470 \sim 550 nm.$$

In the following calculation one can take an average value, e.g.

$$\lambda = 500 nm$$

Furthermore we analyze the proteins are membrane-bound membrane proteins of cells, the diffusion coefficient of the proteins of cell membrane is in the order of magnitude (refer to Fan and Fan [12])

$$D = 10^{-9} \sim 10^{-13} cm^2/s$$

As an average value we take an average value, e.g.

$$D = (3 \sim 5) \times 10^{-10} cm^2/s$$

Substituting above data into (5), one finds that the wave propagation speed of PRET signal is in the range

$$U = (16.8 \sim 28.6) nm/s$$

which is in good agreement to the experimental measurement value

$$U = (18.1 \pm 1.7) nm/s$$

that was measured by Wang et al. [9]. This has examined the efficiency of our physical-mathematical model.

***Two- and three-dimensional analysis.*** For simplicity only the one-dimensional analysis is discussed in the above section. However the two- and three-dimensional analyses have no difficulty and are given in the following.

For two- and three-dimensional cases the equation of motion (1) will be replaced by

$$\frac{\partial C}{\partial t} = D\nabla^2 C \tag{7}$$

in which $C(\mathbf{r},t)$ represents the concentration of proteins of a cell, and $\nabla^2 = \frac{\partial^2}{\partial x^2} + \frac{\partial^2}{\partial y^2}$ for two-dimensional case, and $\nabla^2 = \frac{\partial^2}{\partial x^2} + \frac{\partial^2}{\partial y^2} + \frac{\partial^2}{\partial z^2}$ for three-dimensional case, respectively. The wave solution of equation (7) is

$$C(x,t) = C_0 e^{i(\mathbf{k}\cdot\mathbf{r}-\omega t)} \tag{8}$$

where $C_0$ is a constant, $i = \sqrt{-1}$, $\mathbf{k} = (k_x, k_y, k_z)$ the wave vector, $\mathbf{r} = (x,y,z)$ the radius vector, and $\mathbf{k}\cdot\mathbf{r} = k_x x + k_y y$ for two-dimensional case, and $\mathbf{k}\cdot\mathbf{r} = k_x x + k_y y + k_z z$ for three-dimensional case, respectively. Furthermore the dispersion relation is similar to (3), i.e.,

$$|\mathbf{k}| = \sqrt{k_x^2 + k_y^2 + k_z^2} \equiv k = \frac{1}{\sqrt{2}}\sqrt{\frac{\omega}{D}} \tag{9}$$

The wave velocity formula is also similar to (4), i.e.,

$$\mathbf{U} = \frac{\partial \omega}{\partial \mathbf{k}} \qquad (10)$$

but here the wave velocity $\mathbf{U}$ is a vector rather than a scalar. According to (10) any wave velocity components can be evaluated. Up to now we have not found the experimental data concerning the two- and three-dimensional wave propagation in the mechanical activation in cells, the further calculation cannot be offered yet.

Although the present study is phenomenological, the solution especially the new dispersion relation (3) are very simple and effective, which can serve a mathematical basis to explain the long-range wave propagation induced by external force stimulation to the living cells.

Furthermore there is another example in physics (the phason dynamics of quasicrystals) supporting the diffusion wave theory, we will report the relevant results elsewhere.

**Acknowledgement**  The first author thanks the National Natural Science Foundation of China for financial support. We also thank Dr. S. Chien of the University of California at San Diego and Dr. Y. Wang of University of Illinois at Urban for their helps, so that the figures of Ref [9] with high resolution are available.